\documentclass[nocopyrightspace]{sigplanconf}
\pdfoutput=1

\usepackage{amsmath}
\usepackage{amssymb}
\usepackage{listings}
\usepackage{url}
\usepackage{color}
\usepackage[usenames,dvipsnames]{xcolor}
\usepackage{subcaption}
\usepackage{booktabs}
\usepackage{graphicx}
\DeclareGraphicsExtensions{.pdf}

\begin{document}

\setlength{\pdfpageheight}{\paperheight}
\setlength{\pdfpagewidth}{\paperwidth}

\conferenceinfo{PACT}{'15 San Franscisco, California, USA}
\copyrightyear{20yy} 
\copyrightdata{978-1-nnnn-nnnn-n/yy/mm} 
\doi{nnnnnnn.nnnnnnn}




\titlebanner{banner above paper title}        
\preprintfooter{short description of paper}   

\title{Boosting Java Performance using GPGPUs}
\subtitle{}

\authorinfo{James Clarkson \and Christos Kotselidis \and Gavin Brown \and Mikel Luj\'an}
           {The University of Manchester}
           {\{first.last\}@manchester.ac.uk}

\maketitle
\begin{abstract}
Heterogeneous programming has started becoming the norm in order to achieve better performance by running portions of code on the most appropriate hardware resource.
Currently, significant engineering efforts are undertaken in order to enable existing programming languages to perform heterogeneous execution mainly on GPUs.
In this paper we describe Jacc, an experimental framework which allows developers to program GPGPUs directly from Java.
By using the Jacc framework, developers have the ability to add GPGPU support into their applications with minimal code refactoring.

To simplify the development of GPGPU applications we allow developers to model heterogeneous code using two key abstractions: \textit{tasks}, which encapsulate all the information needed to execute code on a GPGPU; and \textit{task graphs}, which capture the inter-task control-flow of the application.
Using this information the Jacc runtime is able to automatically handle data movement and synchronization between the host and the GPGPU; eliminating the need for explicitly managing disparate memory spaces.

In order to generate highly parallel GPGPU code, Jacc provides developers with the ability to decorate key aspects of their code using annotations.
The compiler, in turn, exploits this information in order to automatically generate code without requiring additional code refactoring.

Finally, we demonstrate the advantages of Jacc, both in terms of programmability and performance, by evaluating it against existing Java frameworks.  
Experimental results show an average performance speedup of 32x and a 4.4x code decrease across eight evaluated benchmarks on a NVIDIA Tesla K20m GPU.
 
\end{abstract}

\section{Introduction}


Heterogeneous programming languages enable developers to execute portions of their code onto specialized devices.
This task, which typically involves offloading work from a \textit{host} onto a more specialized \textit{device} such as a GPGPU, requires a programming model that supports code execution in different \textit{contexts}.
The downside of such programming languages is that they require developers to be aware of the \textit{contexts} when writing code and to ensure that execution and data are synchronized across them.
In this paper we demonstrate that a large amount of this responsibility can be eliminated or handled automatically and, consequently, reduce the burden on the developers.

Current established heterogeneous programming languages, such as CUDA \cite{cuda} and OpenCL \cite{OpenCL}, require developers to logically separate their applications into code that runs either on the host or on the device (known as a \textit{kernel}). 
As a consequence, these approaches require additional code to co-ordinate execution between the host and kernels.

In this paper we demonstrate a simplified heterogeneous programming model, in the context of the Java language, through the use of implicit parallelism and data synchronization. The introduced Java Acceleration system (Jacc) shares many similarities with directive-based approaches like OpenAcc \cite{openacc} and OpenMP 4.0 \cite{openmp_4}.
However, unlike these approaches it also has the ability to automatically optimize execution by eliminating redundant data transfers and executing kernels out of order.
Furthermore, the modular programming interfaces of Jacc enable future extensions that will result in dynamically selecting both the number and types of the underlying devices for code execution (e.g. GPGPUs, FPGAS or ASICs).
In order to achieve that, the only information necessitated from the developer is a model of the heterogeneous portion of the application, captured as a Direct Acyclic Graph (DAG), and the decoration of kernels with metadata that will transparently guide the compilation.
 
Overall the paper makes the following contributions:
\begin{itemize}
\item Introduces an approach to heterogeneous programming that abstracts away low-level hardware details and housekeeping functionalities.
\item Presents the design and implementation of Jacc and all its components.
\item Showcases the efficiency of Jacc by implementing a GPGPU offloading runtime that is capable of handling the asynchronous nature of heterogeneous programs.
\item Provide an in-depth comparative performance analysis of Jacc and standard Java multithreaded benchmarks. The experimental results show that Jacc can provide an order of magnitude improvement in performance while decreasing code size by 4.4x.
\end{itemize}

\section{The Jacc Framework}

\begin{figure}
\centering
\includegraphics[scale=.48,bb=0 0 495 375]{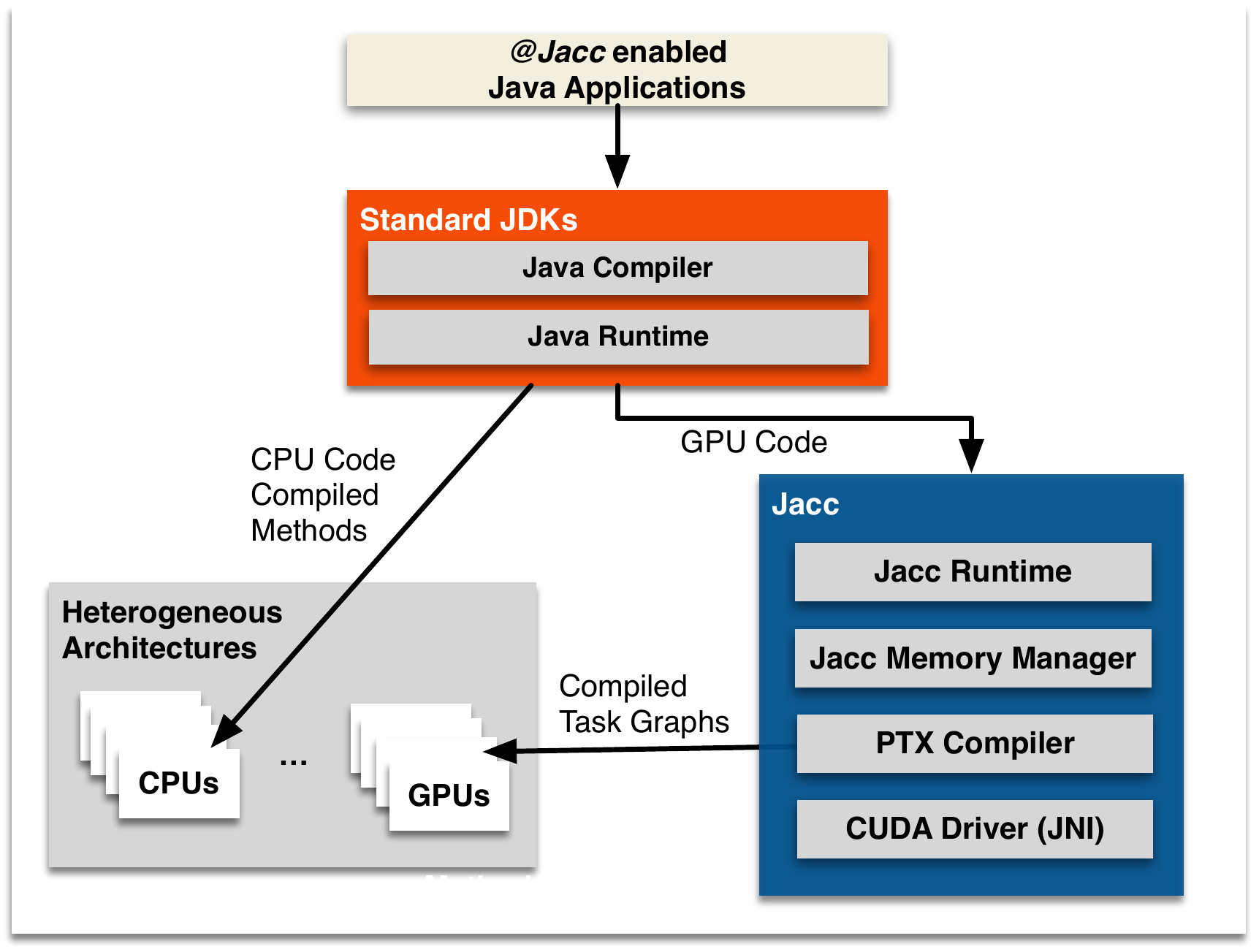}
\caption{Jacc system overview.}
\label{fig:jacc_overview}
\end{figure}

Jacc is a Java based framework that allows developers to program heterogeneous hardware in a simplified manner. In the context of this paper, we showcase Jacc in programming CUDA enabled GPGPUs directly from Java.
As depicted in Figure \ref{fig:jacc_overview}, the two major components of Jacc are: 1) a \textit{just-in-time} (JIT) compiler to compile Java bytecode into low-level machine code for GPGPUs and b) a runtime system designed to manage the execution of application kernels on GPGPUs; all the  compilation, data movement between devices, synchronization and kernel executions are handled automatically by the runtime.

The design philosophy of Jacc is that by providing the right programming abstractions it is possible to create highly-parallel kernels without forcing developers to change their software engineering practices. To that end, Jacc builds on top of two basic blocks: the \textit{task} and the \textit{task graph}.
A task encapsulates all the vital information for executing code in a parallel environment; typically a method reference, a parameter list and some scheduling metadata.
Most importantly, tasks are not restricted to execute on specific hardware. 
They are mapped onto hardware when they are inserted into a task graph; a mapping which can be changed dynamically by the developer.
Tasks can be created from any method in the application; however, methods annotated with the \texttt{@Jacc} annotation create data-parallel kernels using either implicit or explicit parallelism. 
The two different options are described in Section \ref{sec:expressing_parallelism}.

Instead of executing tasks directly on the GPGPU, Jacc follows a more efficient approach and executes them as part of a \textit{task graph} (described in Section \ref{sec:executing_code}).
Executing a single task on a GPGPU requires a number of actions to be performed by the runtime. 
These actions include code compilation, data transfers to the GPU, code execution on the GPU, and data transfers back to the host. 
By using task graphs, Jacc can automatically generate and optimize these actions in a holistic manner. Nodes' re-organization, redundant actions' removal, and early kernel scheduling are some of the optimizations that the usage of task graphs allow.
To model the control-flow between tasks, the task graph is implemented as a \textit{directed acyclic graph} (DAG) which allows Jacc to exploit any available task-parallelism.

Regarding programmability, Jacc exploits many features of the Java language in order to simplify the development workflow. 
Since the code is directly generated from Java bytecode, the need to embed source code inside the application, like OpenCL, or re-parse high-level application source is alleviated.
Moreover, the JIT compiler provides Jacc with the ability to dynamically recompile kernels on-demand which constitutes the a priori assumptions regarding the available resources unnecessary.
Additionally, the dynamic JIT compilation can assist in specializing both compilation and execution for the underlying devices.

Unlike a large number of alternate approaches, such as APAR\-API \cite{aparapi}, Habanero-Java \cite{habanero_java}, JCUDA \cite{jcuda_2009} and Rootbeer \cite{rootbeer}, which generate code via high-level languages, mainly CUDA and OpenCL, Jacc targets the virtual ISA of the GPGPU - PTX \cite{ptx}.

\subsection{Jacc Programmability}

One of the key design features of Jacc is the ability to simplify programming GPGPUs without diverting significantly from standard traditional programming practices of implementing multi-threaded code.
In this section we show how a simple reduction operation, the summation of numbers in an array, can be performed using Jacc in contrast to a vanilla multi-threaded Java implementation. 
In this example, the two main issues faced by the developer are: how to perform the parallel decomposition and how to communicate data between threads.

\subsubsection{Java Implementation}

Listing \ref{lst:java_mt_reduce} shows the Java implementation of the reduction operation.
It decomposes the problem, using a block distribution, in order for the work to be assigned equally across all available processors (lines 16-18).
This produces a two stage algorithm where threads initially reduce their assigned portions of the array individually. After all threads complete the initial step, they reduce their intermediate results in order to produce the final value.
The reduction operation employs atomic operations from the \texttt{java.util.concurrent} library in order to avoid introducing locks and, thus, achieve higher performance. Since there is no direct API support for atomic operations on \texttt{float} primitives they have to be converted into \texttt{int} primitives first (\texttt{AtomicInteger} class as illustrated in lines 24-29).

In order to perform thread assignment, the reduction operation utilizes a thread pool.
Listing \ref{lst:java_mt_executor} shows the code required to: a) create the thread pool (line one), b) create and submit work to the thread pool, and 3) wait for the work to complete (lines 2-15). After the work is submitted the threads execute asynchronously. Consequently, in order to ensure that the reduction operation is completed a \texttt{CyclicBarrier} is used to block execution until all threads have finished\footnote{The barrier is required since the result is only valid once all threads have completed.}.
 
\begin{figure}
\centering
\begin{lstlisting}[caption=Multi-threaded reduction operation in Java,label=lst:java_mt_reduce,numbers=left,frame=tb,backgroundcolor=\color{lightgray}]
public class Reduction
 			implements Runnable {
  final float[] data;
  final AtomicInteger	result;
  final int id;
  final int nThreads;
  final CyclicBarrier	barrier;

  public Reduction(int id,int nThreads, 
	   CyclicBarrier barrier,float[] data,
	   AtomicInteger result) {
					...
	}

  public void run() {
    int work = (data.length + 
				nThreads) / nThreads;
    int start = id * work;
    int end = Math.min(start + work,
 				data.length);
    float sum = 0f;
    for (int i = start; i < end; i++) {
    	sum += data[i];
    }
    int expected = result.get();
    float tmp = 
		Float.intBitsToFloat(expected);
    while(!result.compareAndSet(expected, 
			Float.floatToIntBits(sum + tmp))) {
		expected = result.get();
		tmp = Float.intBitsToFloat(expected);
    }
    barrier.await();
  }
}

\end{lstlisting}

\begin{lstlisting}[caption=Launching work in Java using an ExecutorService,label=lst:java_mt_executor,numbers=left,frame=tb,backgroundcolor=\color{lightgray}]
ExecutorService	executor  = 
	Executors.newFixedThreadPool(threads);
CyclicBarrier barrier = 
	new CyclicBarrier(threads + 1);
barrier.reset();

AtomicInteger result = new AtomicInteger(
	Float.floatToIntBits(0f));
executor.execute(new Reduction(i,threads,
 	barrier, data, result));

barrier.await();
result = 
	Float.intBitsToFloat(result.get());

executor.shutdown();
while (!executor.isTerminated()) {}
\end{lstlisting}
\end{figure}

\subsubsection{Jacc Implementation}

Listing \ref{lst:jacc_reduce_implicit} shows the equivalent implementation of the reduction operation using Jacc.
The immediate observation is that the Jacc implementation requires less lines of code.
This is due to the compiler's ability to support parallelism and atomic operations transparently.
In line three, the \texttt{@Jacc} annotation instructs the compiler that each iteration of the outermost loop (in the first loop-nest) should be assigned to individual threads.
In this example, the iteration space will be defined as $0$ to \texttt{array.length} - the domain of the outermost loop.
The actual number of threads used to execute this code on a GPGPU  is determined by the parameters passed to the runtime system in lines 6 and 7 of Listing \ref{lst:jacc_reduce_taskgraph}.
Here a single thread is started for each element of the array being reduced.
Therefore, \texttt{array.length} threads are used and organised into thread groups of \texttt{BLOCK\_SIZE}.

The \texttt{@Atomic} annotation, used in line one, instructs the compiler that any access made to the \texttt{result} field must be made atomically (using shared memory atomic operations).
The exact atomic operation used can be specified with the \texttt{op} parameter - in this case the compiler will ensure that any assignments to \texttt{result} will be combined with its existing value using addition; effectively turning the assignment into: \texttt{result += sum}.
Unless the field is initialised elsewhere in the task or instructed not to, the compiler automatically initialises atomic variables to zero.

One of the design goals of Jacc is that where possible knowledge of how an application is parallelized is  captured as parameters - via annotations or tasks.
This mean that the underlying Java code still produces a correct result if it is executed in a serial manner; making it possible for a method to remain usable by Java applications and be accelerated by Jacc.
Additionally, this feature of Jacc makes it possible to fallback onto the serial implementation if problems are encountered, such as a device becomes unusable or the compiler is unable to generate GPGPU code.

\begin{figure}
\centering
\begin{lstlisting}[caption=Jacc reduction operation using implicit parallelism,label=lst:jacc_reduce_implicit,numbers=left,frame=tb,backgroundcolor=\color{lightgray}]
@Atomic(op = ADD) float	result;

@Jacc(iterationSpace = ONE_DIMENSION)
public void reduction(@Read float[]
 							array) {
	float sum = 0;
	for(int i = 0; i < array.length;++) {
		sum += array[i];
	}
	result = sum;
}
\end{lstlisting}

\begin{lstlisting}[caption=Launching work in Jacc using the TaskGraph,label=lst:jacc_reduce_taskgraph,numbers=left,frame=tb,backgroundcolor=\color{lightgray}]
DeviceContext gpgpu	=
  Cuda.getDevice(0).createDeviceContext();

Task task = Task.create(
		Reduction.class, methodName,
		new Dims(array.length), 
		new Dims(BLOCK_SIZE));

task.setParameters(r, data);
tasks = new NewTaskGraph() {
	@Override
	public void create() {
		executeTaskOn(task, gpgpu);
	}
}	
tasks.execute();
\end{lstlisting}
\end{figure}

\begin{figure*}[t]
\centering
\includegraphics[scale=.5,bb= 0 0 812 304]{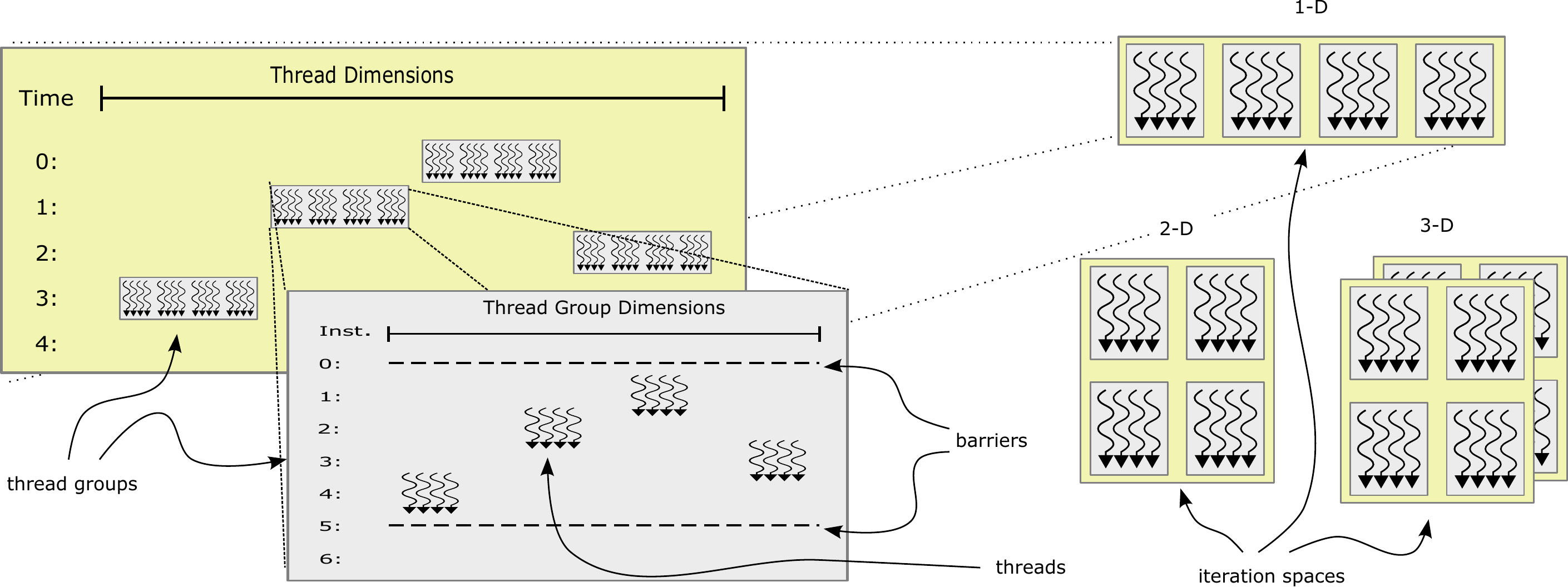}
\caption{Fine-grained execution model}
\label{fig:execution_model}
\end{figure*}

Listing \ref{lst:jacc_reduce_taskgraph}, shows how this code is configured for GPGPU execution.
The first step is to create a new task, lines 3-6, with lines 6-7, defining how the iteration space is mapped onto individual threads.
Although this code simply maps each iteration onto a unique thread,it is also possible to adjust the amount of work assigned to each thread by specifying less threads than points in the iteration space.
For example, line 6 can be changed to start \texttt{array.length / BLOCK\_SIZE} threads - this will force the GPGPU to execute the kernel using a block cyclic mapping.
This is useful in the reduction example, as it can be used to reduce the number of threads competing to perform atomic operations.

Finally, to map the task onto a GPGPU for execution, a new task graph has to be created as shown in lines 10-15.
The method \texttt{executeTaskOn} inserts a new node into the task graph and maps the execution of this node onto the specified GPGPU.
Once built, the task graph can be executed by invoking its \texttt{execute} method.
This method blocks until either all tasks in the task graph have completed or an exception occurs.
If all tasks complete successfully, Jacc ensures that all outstanding updates to the host memory are visible before \texttt{execute} completes.

The benefits of Jacc become clear when comparing Listings \ref{lst:java_mt_reduce} and \ref{lst:java_mt_executor} with \ref{lst:jacc_reduce_implicit} and \ref{lst:jacc_reduce_taskgraph} respectively. 
The code is more concise and requires 45\% fewer lines of code in this particular example.
Another major advantage of Jacc is that tasks can be created from any method, eliminating the need to create new classes for each new parallel kernel.
This property helps to increase code re-user and also maintains backwards compatibility with applications that do not use the Jacc framework.

\subsection{Jacc Language Design}

One of the design principles of Jacc is to produce high performance GPGPU code without having to re-write applications in a device specific manner.
Therefore, instead of developing a new superset of the Java language, existing language features are used in order to provide the framework with extra information about the application; this is done by providing metadata to Jacc through the task abstraction and the use of annotations.

This section presents the design and programmability aspects of Jacc.
First, the parallel execution model (Section \ref{sec:execution_model}) and how it relates to the Java Memory Model (Section \ref{sec:java_memory_model}) is explained. In turn, the programming principles of Jacc are explained.

\subsubsection{Parallel Execution Model}
\label{sec:execution_model}

One of the key distinctions between traditional processors and GPGPUs is that GPGPUs are optimized for throughput over latency.
Therefore, efficient GPGPU execution is achieved by splitting the work into large numbers of small threads, potentially tens of thousands to millions, opposed to traditional processors which prefer smaller numbers of larger threads. 
It is their ability to keep large numbers of threads in flight that allows GPGPUs to hide memory latency better than latency optimised processors.

Since the Java language was not designed with GPGPU execution in mind, a different execution model is needed; one which enables code to be executed by large numbers of threads. 
Therefore, in order to generate the abundance of threads required to sustain throughput on the GPGPU, Jacc adopts a fine-grained parallel execution model. By using this model, it is possible for each iteration in a loop to be assigned to its own individual thread - since we assume thread management has a very low-overhead.

Figure \ref{fig:execution_model} depicts how the parallel execution model works.
In Jacc's execution model, the developer simply maps each point of a problem's \textit{iteration space} onto a thread. 
As the iteration space may be large, it is divided into a number of equal sized units called \textit{thread groups}; adjusting the group size can assist in exploiting a problem's locality in the same way as blocking. 
These groups are used as the basic unit of scheduling on the GPGPU.

Finally, a key point is that thread groups can be executed in \textit{any} order meaning that the only form of global inter-thread communication is via memory (data independent tasks). Nevertheless, no guarantees of ordering between actions performed by different thread groups is provided.
In order to provide memory ordering guarantees between threads, synchronization barriers between all threads within the same thread group have to be inserted.

\subsubsection{Memory Synchronization}
\label{sec:java_memory_model}

The Java Memory Model (JMM), which is defined in the Java Language Specification (JLS) \cite{jls}, requires serial consistency of all memory operations (memory accesses are made visible in program order). In order to achieve that in shared-memory multi-threaded applications, synchronization mechanisms are employed. The standard synchronization mechanisms of Java (synchronization primitives, synchronized and atomic data structures), are inadequate for GPGPU execution because they are too coarse.

For instance, the use of \texttt{synchronized} fields may result in serializing the memory access of thousands of GPU threads.
This will, undoubtedly, cause performance degradation on the GPGPU because it will need to handle the divergence of all the threads.
Therefore, in order to allow synchronization at a much finer granularity the GPU's memory barriers are exposed at the programming level which allows synchronization between threads within the same thread group\footnote{Note that the \texttt{synchronized} keyword is ignored on the GPGPU.}.

Additionally, Jacc provides two additional ordering guarantees. Firstly, the ordering inside the task graph is preserved on the GPGPU  - this means that all changes performed by one task are visible to any subsequent tasks that use the same data only on the device. Secondly, the task graph executes atomically and all memory updates are made visible to the host before the task graph completes. However, it should be noted that in the case where primitives are communicated between the host and the GPGPU, the use of the \verb+volatile+ keyword may be required in order value caching by the compiler.

\subsubsection{Annotations}

Jacc utilizes heavily Java annotations in order to both tag methods for GPU compilation and to instruct the Jacc compiler on how to optimize the compiled code.
Table \ref{tbl:annotations} provides an overview of the different annotations, their parameters and their use within the Jacc framework.
This is a similar approach to directive-based programming languages such as OpenMP \cite{openmp_4}, HMPP \cite{hmpp}, and OpenAcc \cite{openacc}.

\begin{table*}
   \centering
   \begin{tabular}{@{} cccp{3cm}p{7.5cm} @{}} 
      \toprule
      Annotation & Target & Parameter & Options & Description\\
      \midrule
      \verb+@Jacc+     & method & \verb+iterationSpace+ & \verb+NONE+, \verb+ONE_DIMENSION+, \verb+TWO_DIMENSION+, \verb+THREE_DIMENSION+  & Defines the iteration space to use in parallelizing the method.  \\
				 & & \verb+exceptions+ & \verb+true+, \verb+false+ & Defines whether the compiler should insert exception checks into the kernel. \\
	\verb+@Atomic+ & field & \verb+op+			& \verb+NONE+, \verb+ADD+, \verb+SUB+, \verb+AND+, \verb+OR+, \verb+XOR+  & Declares that this field should only be updated using the specified shared memory atomic operation. (If none is specified the compiler will infer the operation from the code.)\\
	\verb+@Shared+ & field & & & Each thread group should share a copy of this field.\\
	\verb+@Private+ & field & & & Each thread should have a private copy of this field.\\
	\verb+@Read+ & parameter & \verb+cachable+ & \verb+true+, \verb+false+ & Declares that this is a read-only parameter by the kernel.\\
	\verb+@Write+ & parameter& \verb+cachable+ & \verb+true+, \verb+false+ & Declares that this is a write-only parameter by the kernel. \\
	\verb+@ReadWrite+ & parameter & \verb+cachable+ & \verb+true+, \verb+false+ & Declares that this a read/write parameter by the kernel. \\
      \bottomrule
   \end{tabular}
   \caption{Details of the available annotations in Jacc.}
   \label{tbl:annotations}
\end{table*}

\subsubsection{Expressing Parallelism}
\label{sec:expressing_parallelism}

Generally, there are two ways in which parallelism can be expressed in parallel languages: explicitly or implicitly.
Although the Jacc framework supports both, implicit parallelism is strongly encouraged since the application code remains unchanged allowing the development of fallback mechanisms in case errors are encountered.

Regarding \textit{implicit parallelism}, the Jacc compiler has the capability to parallelize certain classes of loop-nests.
This is performed by re-writing loops in order to assign iterations to individual threads.
The loops are processed in order, starting with the outermost and moving towards the innermost.
The number of loops that are re-written is controlled by the \verb+iterationSpace+ parameter of the \verb+@Jacc+ annotation.
For example, the reduction operation in Listing \ref{lst:jacc_reduce_implicit} will be re-written by the compiler (using methods from the Jacc \verb+Helper+ library) resulting in Listing \ref{lst:jacc_reduce_implicit_internal}.

\begin{figure}
\centering
\begin{lstlisting}[caption=Compiler generated implementation ,label=lst:jacc_reduce_implicit_internal,numbers=left,frame=tb,backgroundcolor=\color{lightgray}]
@Jacc(iterationSpace = NONE)
public void reduction(@Read float[] 
                              array) {
	float sum = 0;
	int tid = Helper.getTidX();
	int numThreads = Helper.getNThreadsX();
	for(int i=tid; i<array.length;
                    i+=numThreads){
		sum += array[i];
	}
	result = sum;
}
\end{lstlisting}
\end{figure}

The advantage of this approach is that the original method can still be executed correctly, since the compiler is able to transform the code automatically, by ignoring the annotation.

Regarding \textit{explicit parallelism}, in cases where it is not possible to express a kernel using a single loop-nest, Jacc provides the developer with two choices: a) to split functionality across multiple kernels or, b) to manually parallelize the code.
If a developer wishes to manually parallelize code then it is performed in the same way as CUDA and OpenCL - via a set of built-in functions.
The advantage of this approach is that developers can create highly optimized parallel code for a specific device.
Unfortunately, this comes at the expense of reduced code re-use as Java applications cannot readily use this code.

\subsubsection{Communications}

One common issue when writing parallel code is communicating data between threads.
In Jacc, all communication is performed via memory.
Nevertheless, in order to optimize some types of communication, Jacc supports a range of functionality that allows data to be communicated:

\begin{itemize}
\item Inter-thread Communication: A typical problem encountered when parallelizing certain kernels is communicating data between threads. Normally, this communication is performed via memory. To avoid having to serialize access to shared variables, Jacc supports atomic operations on shared memory.
As atomic operations operate on a memory location it is only possible to generate atomic accesses for operations on fields and arrays; not local variables.

\item Inter-method Communication: More complex applications may require multiple methods to be executed on the hardware accelerator.
In these cases it is likely that a method execution may depend on the results of the execution of a prior method.
\end{itemize}

\subsection{Executing Tasks}
\label{sec:executing_code}

In order for Jacc to execute tasks, it is required from the developer to pass a DAG (or a task graph using Jacc terminology) to the runtime system.
The task graph models the control-flow of all interactions between the host and the GPGPU.
From the provided task graph, the runtime system applies a lowering process where each task is decomposed into a series of lower-level tasks. Code compilation, data transfers and synchronization barriers are examples of these lower-level tasks.
Using the information contained within the task graph it is possible to infer all the data dependencies between tasks - allowing nodes to be re-organized whilst satisfying these dependencies. Once lowered, the runtime system traverses the task graph looking for opportunities to eliminate, merge and re-organize these nodes.
During execution, the runtime system simply traverses the optimized task graph and executes each node it encounters.

One of the main motivations for using a DAG based approach is that it allows the clean separation of \textit{which} code is executed from \textit{where} it is executed.
This provides developers with the ability to co-locate kernels with regular Java code instead of forcing them re-factor their code and place all kernels into a single class. 
Since all DAG nodes can be parameterized, the introduced techniques will also work for complex multi-kernel, multi-device or mixed-device applications in the future.


\section{Jacc Internals}

\subsection{Compiler Design}

One of the beneficial features of managed languages, and Java in particular, is their platform independence via their bytecode representation. The application source code is initially compiled into an \textit{architecture-neutral} intermediate representation that allows efficient generation of machine code at runtime; typically this is the role of a dynamic JIT compiler.
By constructing our compiler to work at the bytecode level, we avoid the need to parse source-code which makes the compilation process more efficient.
Additionally, by developing a JIT compiler we gain the following advantages: 1) we do not require a priori knowledge of the target platform since the runtime system can discover this information automatically, and 2) as we know exactly what hardware is available the compiler can be designed to be overly aggressive and customize code generation for a particular architecture as opposed to static compilers which are generally more conservative.

\begin{figure}[t]
\centering
\includegraphics[width=.45\textwidth,bb=1 2 549 460]{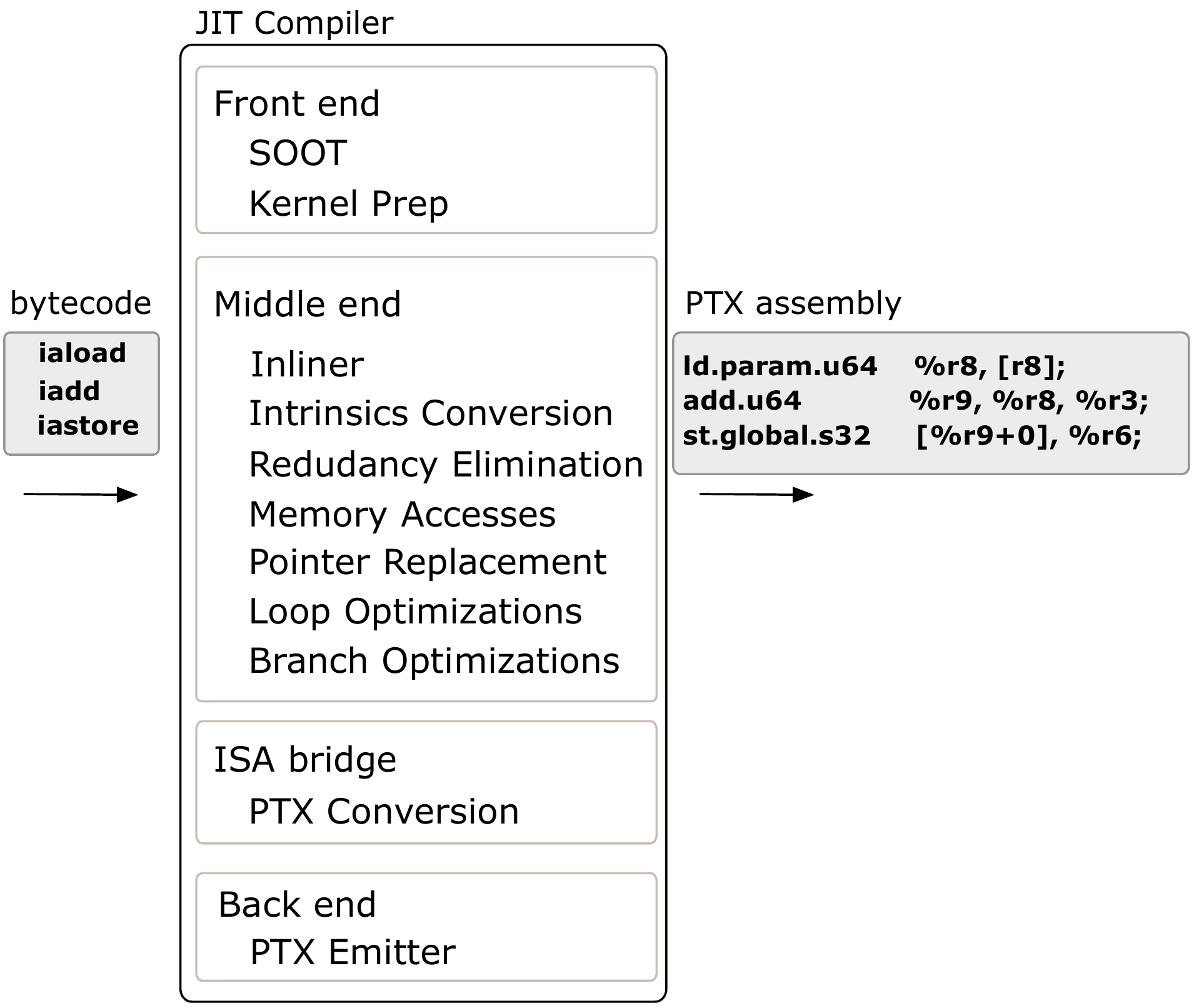}
\caption{Compiler Structure}
\label{fig:compiler_arch}
\end{figure}

Figure \ref{fig:compiler_arch} illustrates the structure of the JIT compiler used in Jacc.
The compiler, which generates PTX code directly from Java bytecode, is organized in three layers: the front-end - responsible for parsing bytecode; the mid-end - responsible for transforming and optimizing the code for a GPGPU; the back-end - responsible for emitting the low-level machine-code. 
Although, out current implementation of Jacc targets PTX, we have constructed the compiler to be extensible to allow it to generate code for a range of heterogeneous devices.
For example, we are looking at integrating support for generating HSAIL \cite{hsail} to target HSA (Heterogeneous System Architecture) \cite{hsa} enabled devices and OpenCL SPIR \cite{spir} to target OpenCL enabled devices.

The front-end of the Jacc compiler, currently relies on the SOOT framework \cite{soot_1999}, for parsing bytecode and providing a high-level IR; hence, all our optimizations are performed using SOOT's JIMPLE IR.
The advantage of doing this is that we are able to leverage many of the existing SOOT analyses to help produce good quality of generated code.

To avoid modifying the underlying bytecode, when compilation is requested, a new class is created which holds a copy of the method to be compiled. During the kernel preparation stage, the new class is created from a template that defines device specific behavior, such as entry points, exception handlers and pointers to different memory spaces.

Once the new class is ready, the compiler can optionally parallelize the code.
This transformation searches for the first loop-nest and rewrites the updated schedule of the induction variable in one or more loops; always starting with the outer-most loop.
The update is made depending on the value of the \texttt{iterationSpace} parameter in the code.
Despite being a crude technique, it allows a large number of kernels to be automatically parallelized.
In our experience, the majority of kernels that we could not auto-parallelize was due to the fact that they contained multiple loop-nests.

Presently, we only support the parallelization of the first loop-nest encountered in a kernel.
The reason for this is that the \texttt{Jacc} annotations can only be applied at a method-level of granularity in Java 7.
If annotations can be applied to loop-nests, in a similar style to OpenMP, then multiple loop-nests can be supported.

Once the kernel has been parallelized, the mid-end, which has been specifically designed to generate high quality GPGPU code, starts code generation.
In general, this involves optimizing the control-flow of the kernel, eliminating redundant computation and utilizing the features of the hardware correctly.
The second responsibility of the mid-end is to generate code to target device specific functionality.
This is typically in the form of employing device specific special instructions.
This is achieved by using compiler intrinsics that replace method calls with low-level instructions, such as \texttt{sin} and the memory accesses/pointer conversion, generating efficient load/store instructions.

A common issue involved in mapping our high-level IR onto machine instructions, is that some JIMPLE statements produce invalid assembly statements.
Therefore, before performing code generation we run the JIMPLE IR though an ISA bridge which rewrites any incompatible statements.
For example, one issue is that PTX requires function parameters to be passed via registers.
Hence, in order to pass constants into functions we must assign their values into locals first.
One all incompatible statements are removed from the IR, it is then passed to the PTX emitter which is responsible for converting each statement in the IR into a corresponding PTX statement.

\subsubsection{Optimizing Control-Flow}

One of the major sources of performance losses on GPGPU architectures is control-flow divergence.
This is normally caused by the presence of branches and function calls in the code.
The most severe slowdowns are observed when all threads in a group take a different path through the code.
Nevertheless, there is still a large performance penalty even if all threads in a group branch in a uniform manner.

To reduce the amount of branching inside a kernel, the compiler attempts to inline all method calls and optimize the layout of loops and branches. Consequently, the inliner removes all function calls and the control-flow optimizations eliminate any unnecessary branches from the kernel. If branches that cannot be eliminated are encountered, we have found that it is generally better to replace branches to condition statements with a duplicate of the statement.
Additionally, the Jacc compiler attempts to fully exploit the fact that PTX supports predicated execution by replacing simple branch statements with predicated instructions where possible.

\subsubsection{Eliminating Redundant Computation}

Eliminating redundant computations is of paramount importance for performance improvements in the context of GPGPU execution. 
To achieve this, a set of standard optimization phases are employed by the SOOT framework. Examples of such optimizations are: common subexpression elimination, loop invariant code motion, copy propagation, constant folding, straightening and dead code elimination.

\subsection{Runtime System}

\subsubsection{Memory Management}

One of the key components of the runtime system is the memory manager. Each device has it own manager that provides a mechanism for the runtime system to transfer data to and from the different types of memory available on the device. Additionally, the memory manager is responsible for informing the compiler on how to layout the data.
This is achieved through the use of a \textit{data schema} which maps each element of a composite type into a specific memory location; normally this is relative to a given address. If the runtime system wishes to transfer data to a device it must serialize each object according to the schema in the memory manager.

A key feature of the memory manager is its ability to handle persistent state on each device. This allows data to stay resident on the device across multiple kernel executions eliminating the need to constantly copy data between the host and device.

We ensure shared state remains consistent by requiring that the TaskGraph executes atomically (i.e. no modifications are allowed to the objects on the host while the TaskGraph is executing.). Once the TaskGraph completes the runtime system will ensure all object state is synchronized between the host and the device.

\subsubsection{Data Serialization}
\label{sec:serialisation}

In order for any kernel to execute correctly, the memory manager must pre-load the device with data in a suitable form.
Generally, variables or arrays of primitive types can be copied "as-is" and composite types are laid out according to a data schema.
In the simplest case, the data schema does nothing more than turn an object into a C-like \texttt{struct}.
However, problems occur when objects contain references to other objects, as this requires all referenced objects to be available on the device.

In order to tackle the data serialization process of objects, we experimented with a few alternatives. Initially, we followed a deep copy approach that lead to performance problems. We discovered that while object graphs can become significantly large, in particular when complex codes are utilized, only a small fraction of them is actually utilized by the kernel. That resulted in the unnecessary and costly, in performance terms, transfer of a large number of unused objects.

Consequently, we developed a novel compiler driven approach that builds data schemas on-demand during compilation.
During the compilation process, when new composite-types are discovered, the compiler requests data schemas for them from the memory manager.
If such schemas do not exist, new ones are created on-demand. However, those schemas contain only the elements defined in the top-level of the composite-type - i.e. no inherited fields.

As the compiler progresses, it tracks which fields are accessed and modified by the kernel and records this information in the data schema. The data serializer exploits this information to determine which variables should be transferred to and from the device.
This means that while space is allocated for variables, it is only populated if they are fields are actually used; if a variable is not used it is not transferred. 

Although the compiler constructed schemas enable the execution of a number of kernels, they also have a number of drawbacks.
The most prominent is that this technique causes difficulties when objects access inherited fields - as this requires the compiler to allocating space to accommodate the fields of each object and all fields declared by each super-class.
One possible solution for this is to partially compile kernels to capture all accesses/modifications made to objects and feed this information into a memory optimization phase.
This would enable object layouts to be optimized for streaming access and eliminate redundant fields.

\subsection{Java Language Features}

In order to execute even the simplest kernels, the compiler needs to implement support for advanced features in the Java language.
For example, the kernel may be either a static or an instance method, it may access a class variable or allocate new objects.

\subsubsection{Objects}

One of Jacc's current limitations is that does not interact directly with the garbage collector of the JVM.
This means, in general, that we only support the manipulation of existing objects on the GPGPU.
The only exception to this rule is when created objects cannot escape; these objects can be statically allocated or optimized away by the compiler.
Although this sounds like a major limitation, we have found in practice that most tasks that are amenable for GPGPU offloading perform some form of volume reduction, like our reduction example, and object creation is often not needed.  

Objects are created according to a \textit{data schema} that specifies how the fields should be laid out in memory (discussed in \ref{sec:serialisation}).
We simply treat objects like C-like \texttt{structs} where fields are located at a fixed offset from the start of an object.
To handle inheritance, we insert all fields declared by superclasses into the same struct, as this avoids using indirection which incurs a performance penalty on GPGPUs.
The majority of objects are created by the serializer when task parameters are copied to the device before execution.
There are a few exceptions where the compiler is able to statically determine the space requirements for a new object but dynamic object allocation is not supported.


At present we do not maintain object headers, typically to to reduce storage requirements and improve serialization times.
The consequence is that we do not yet support reflection or the \texttt{instanceof} keyword.
However, there is no technical reason why support can not be added at a later date.

To support placing objects in the different memory spaces on the GPGPU, we tag each object with metadata that allows us to determine which memory space it is located in.
For the most part this is transparent to the developer, however, Jacc provides the ability to specify which memory space a variable should reside by annotating fields with the \texttt{Shared}, \texttt{Private} or \texttt{Constant} annotations.

\subsubsection{Virtual and Static Method Calls}

While the compiler supports method calls, in practice all method calls are removed because of the compiler's aggressive use of inlining. 
Currently, the only cases that method calls are unsupported are when the compiler can not resolve the method being invoked statically; typically, this occurs when developers use Java interfaces.
To remedy this situation in the future we plan to allow the compiler to glean type information from a tasks parameter list.
 
The kernel itself is able to be either a static or a virtual method.
The only difference between the two, from the compiler's perspective, is that it must insert the \texttt{this} object reference as the first parameter passed to the method.
The advantage of virtual methods is that the \texttt{this} object reference neatly encapsulates state that needs to be shared among multiple kernels.

\subsection{Memory Allocations}

Jacc is able to support the \texttt{new} keyword under certain circumstances.
The Jacc compiler will try to inline the constructor and if any memory is required it can allocate it as a stack variable.
Additionally, the use of inlining means we can eliminate a number of field accesses using scalar replacement.
If the developer wishes to allocate memory in a certain memory space, the variable must be declared as a field with the declaration being decorated with an annotation specifying the memory space.

\section{Evaluation}
\label{sec:evaluation}

\begin{figure*}[tbh]
\begin{subfigure}[b]{0.48\textwidth}
\centering
\includegraphics[scale=.48,bb=7 20 477 271]{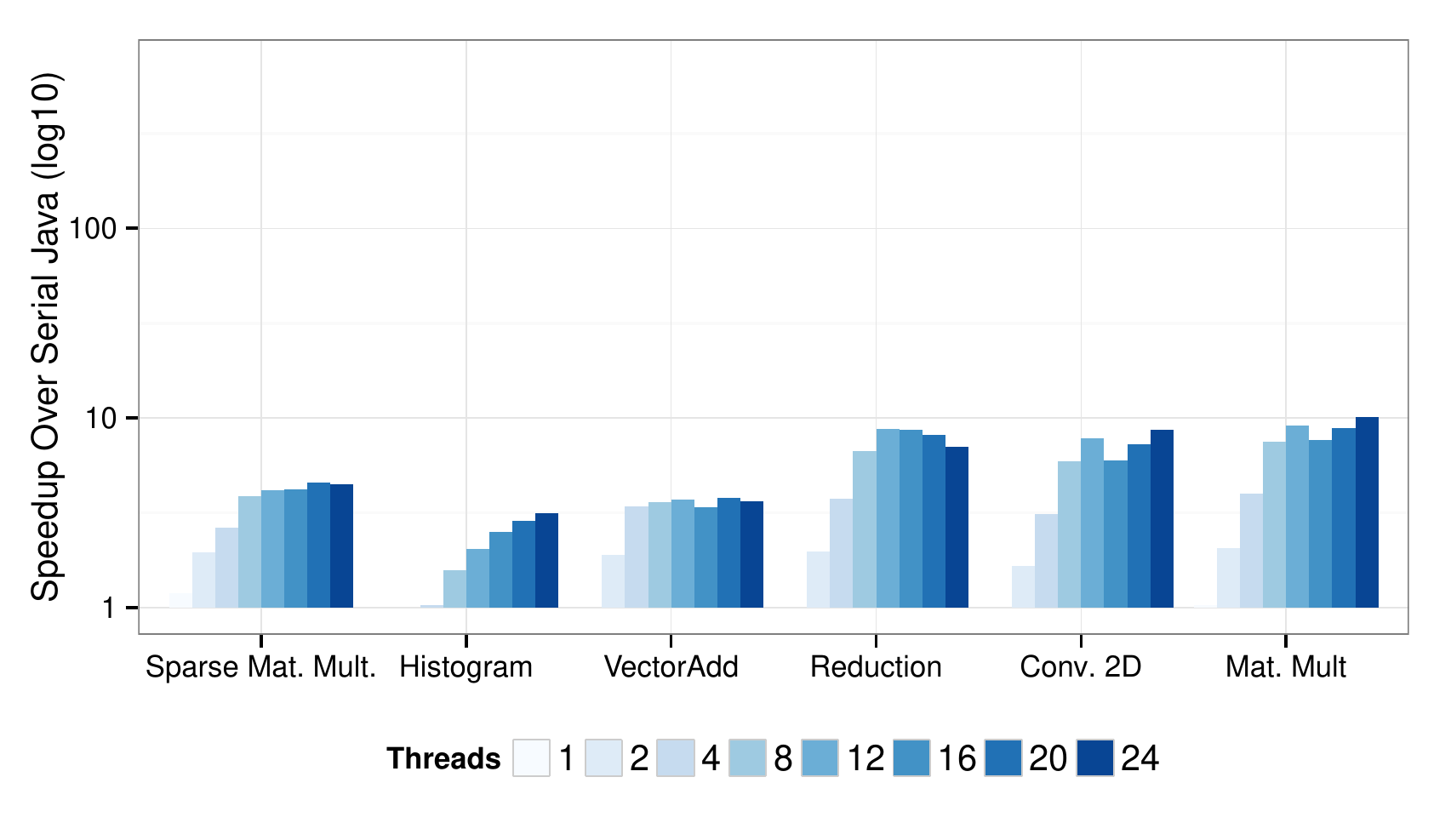}
\caption{}
\label{fig:mt_scaling}
\end{subfigure}
~
\begin{subfigure}[b]{0.48\textwidth}
\centering
\includegraphics[scale=.48,bb=7 20 477 271]{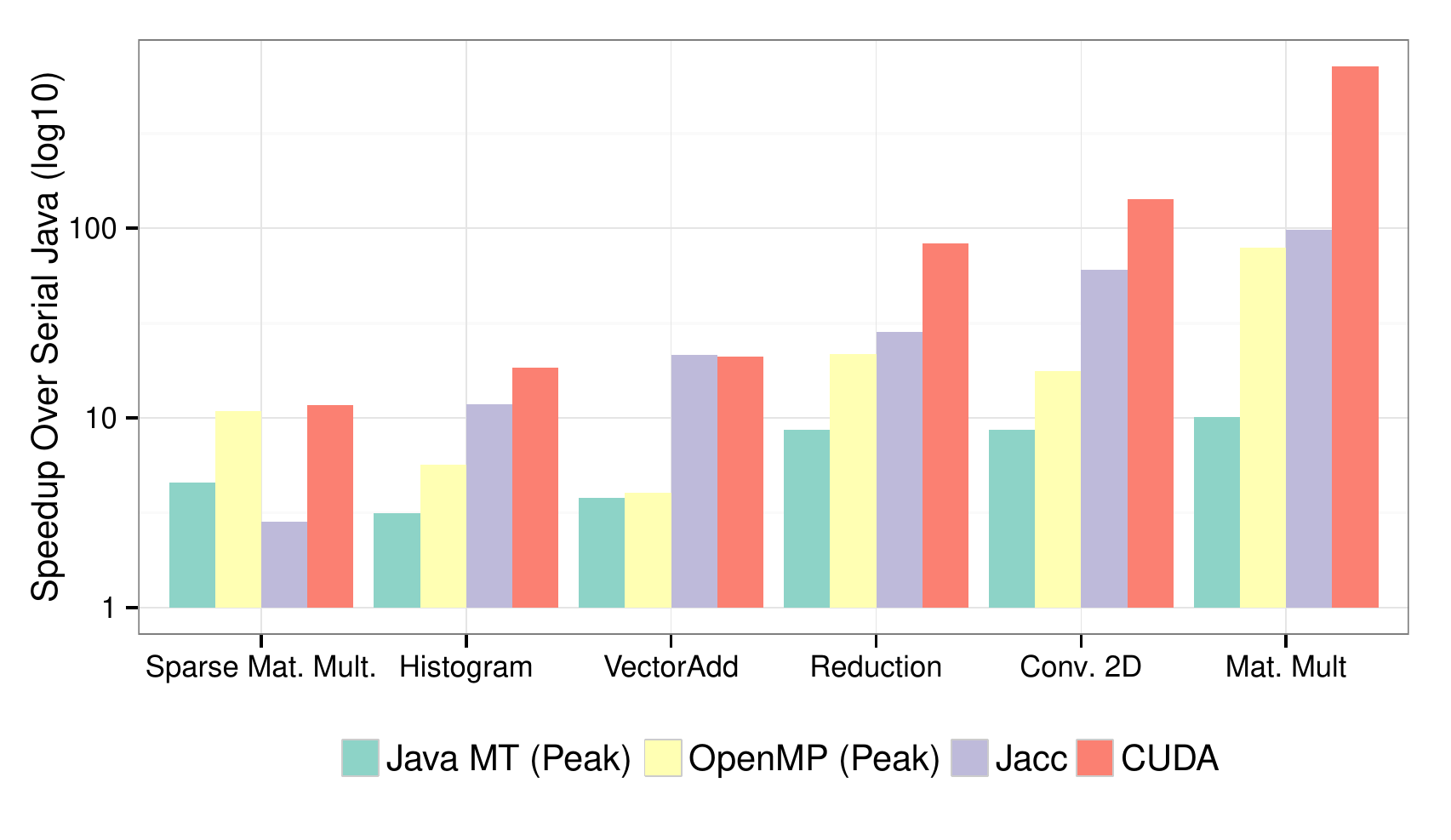}
\caption{}
\label{fig:jacc_benchmarks}
\end{subfigure}
\caption{Homogeneous scaling versus heterogeneous acceleration. (Left) The speedups obtained using multi-threaded Java code. (Right) The speedups obtained using GPGPU acceleration.}
\end{figure*}

In order to assess Jacc's performance we perform various comparisons against: 1) serial Java implementations , 2) multi-threaded Java implementations, 3) OpenMP and CUDA implementations, and 4) against the more mature APARAPI \cite{aparapi} framework, that uses OpenCL \cite{OpenCL} to generate high quality code for a range of heterogeneous devices.

\subsection{Hardware Platform}

The experimental hardware platform has two Intel Xeon E5-2620 processors (12 cores / 24 threads total @2.0 GHz), 32GB of RAM and a NVIDIA Tesla K20m GPU with 5 GB of memory. Regarding the experimental software stack, CentOS 6.5, CUDA 6.5 and  Java SDK 1.7.0\_25 were used. 

\subsection{Benchmarks}
\label{sec:benchmarks}

The evaluated benchmarks represent a number of key numerical kernels that are commonly found in benchmark suites, such as the NASA Parallel Benchmark Suite \cite{npb}, Parsec \cite{parsec}, Rodinia \cite{rodinia} and Linpack \cite{linpack}.
These benchmarks are designed to measure the quality of the  generated code and the efficiency of the runtime system running under ``ideal conditions" (as explained in Section \ref{idealperf}).
The benchmarks used for the performance evaluation are\footnote{All CUDA implementation are taken from the CUDA SDK unless otherwise specified.}:
\begin{description}
\item [Vector Addition] performs the addition of two 16,777,216 element vectors.
The times are measured over 300 iterations of the benchmark.

\item [Reduction] performs a summation across an array of 33,554,432 elements with the time being measured over 500 iterations of the benchmark.

\item [Histogram] produces frequency counts for 16,777,216 values, placing the results into 256 distinct bins. 
The time is measured over 400 iterations of the benchmark.

\item [Dense Matrix Multiplication] performs a multiplication of two $1024 \times 1024$ matrices.
The time is measured over 50 iterations of the benchmark. Furthermore, the OpenMP implementation uses the OS supplied \verb+libatlas+ library and the CUDA implementation uses the vendor supplied \verb+cuBLAS+ library.

\item [Sparse Matrix Vector Multiplication] performs a sparse matrix-vector multiplication using a $44609 \times 44609$ matrix with $1029655$ non-zeros (The bcsstk32 matrix from Matrix Market). The time is measured over 1400 iterations of the benchmark. The CUDA implementation uses the vendor supplied \verb+cuSPARSE+ library.

\item [2D Convolution] convolves a $2048 \times 2048$ image with a $5 \times 5$ filter.
The time is measured over 300 iterations of the benchmark.

\item [Black Scholes] is an implementation of the Black Scholes \cite{blackscholes} option pricing model. The benchmark is executed to calculate 16,777,216 options over 300 iterations and is supplied as an example in the APARAPI source code.

\item [Correlation Matrix] is an implementation of the Lucene \cite{lucene} OpenBitSet ``intersection count''. The benchmark is executed using 1024 Terms and 16384 Documents and is supplied as an example in the APARAPI source code. Only a single iteration is performed.

\end{description}

\subsection{Measuring Performance and Programmability}
\label{idealperf}
The performance of each benchmark is calculated by measuring the time to perform the specified number of iterations of the performance critical section of the benchmark.
Each quoted performance number is an average across a minimum of ten different experiments. 
To ensure that we accurately measure performance on the GPGPU we tried to minimize the amount of overheads incurred.
The reported Jacc execution times are inclusive of a single data transfer to the device and a single transfer to the host but exclusive of JIT compilation times. 
This is done in order to measure the peak-performance of Jacc generated code.

To assess the impact on programmability, we take the stance that code complexity is proportional to code size. 
With that invariant, we demonstrate that applications can be accelerated, using GPGPUs, without requiring an excessive amount of code. We assess this by measuring the number of source code lines required for each implementation. More specifically, only the code that is used to express the parallel kernels is taken into consideration. Other code segments that regard thread or device setup activities are not accounted. This is done because the supporting code is constant between kernels and is of similar length between Jacc and vanilla Java multi-threaded implementations.


\subsection{Java Multi-Threaded Performance}

Figure \ref{fig:mt_scaling} shows the speedups achieved by converting from serial to multi-threaded Java implementations.
These results show that these benchmarks scale with increased thread counts.
That the biggest improvements in performance are achieved when at most one thread is present on each physical core, while the thread count is less than 12.

Finally, the strong divergence from ideal scaling means that even in the hypothetical scenario of achieving near-perfect results the number of cores/processors required to match GPGPU performance would be significantly high.

In order to strengthen the confidence of our results (i.e. folding away possible inefficient Java implementations) we have also implemented all benchmarks in OpenMP 3.2.
Figure \ref{fig:jacc_benchmarks} shows that, with the exception of the sparse matrix vector multiplication benchmark, Jacc still outperforms the OpenMP implementations.

Furthermore, in order to provide a highly optimized OpenMP version the SGEMM implementation from \texttt{libatlas} for matrix multiplication has been used.
Results indicate that even in this case Jacc is still able to outperform OpenMP, albeit by a reduced margin in comparison to Java multithreaded implementations.


\subsection{Heterogeneous Performance}

\begin{figure*}[tbh]
\begin{subfigure}[b]{0.45\textwidth}
\centering
\includegraphics[scale=.45,bb=7 20 477 271]{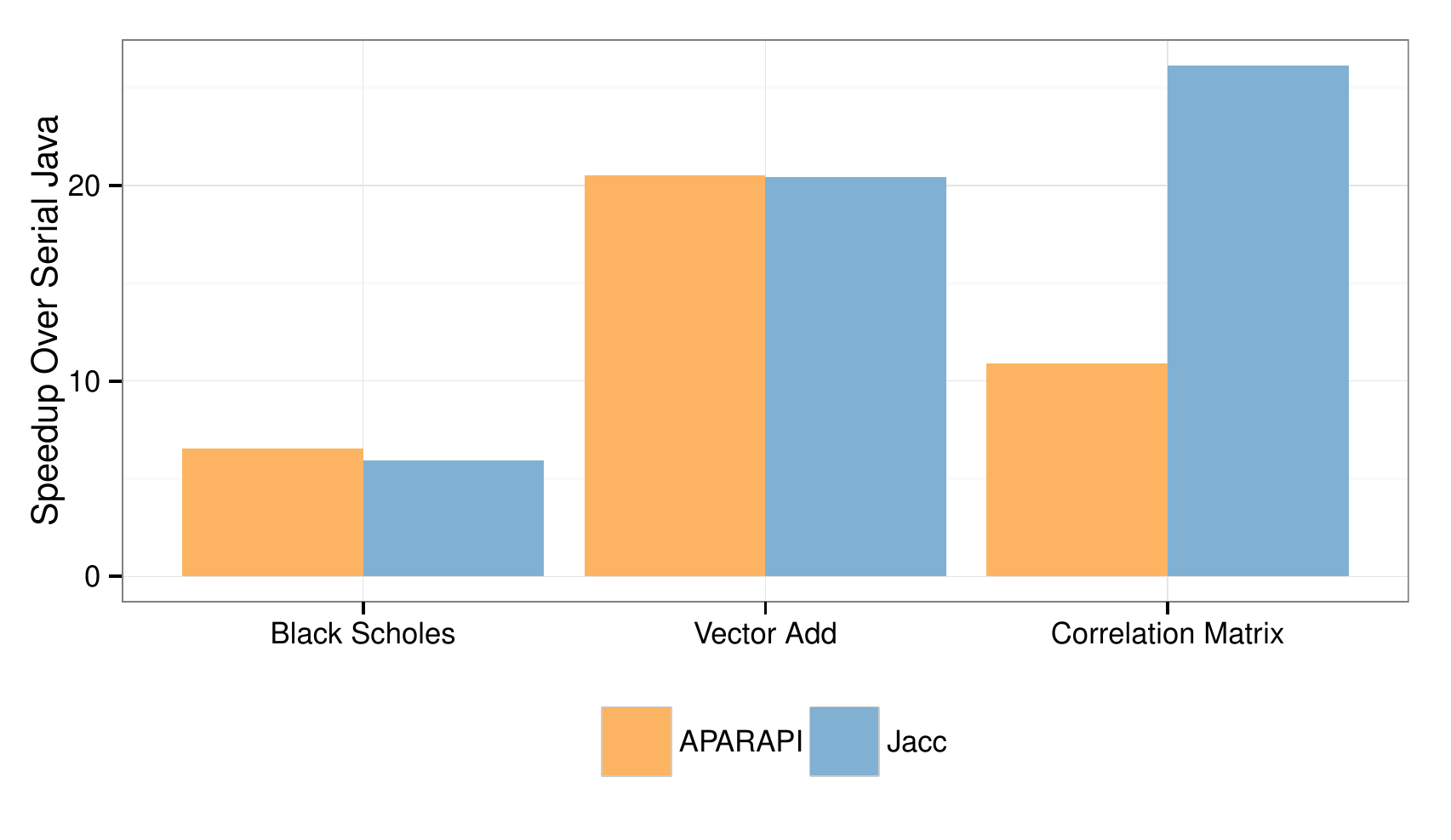}
\caption{}
\label{fig:aparapi_comparison}
\end{subfigure}
~
\begin{subtable}[b]{0.54\textwidth}
	\small
\centering
\begin{tabular}{lrrcccc}
  \toprule
 & \multicolumn{3}{c}{Speedup} &  \multicolumn{3}{c}{Lines of Code} \\
  \cmidrule(r){2-4}  \cmidrule(r){5-7} 
Benchmark & Serial & \multicolumn{2}{c}{Java MT Peak}  & Java MT & Jacc & Reduction \\ 
  \midrule
Vector Add & 21.52 & 6.00 & (20) & 40 & 6 & 6.67x \\ 
  Matrix Mult. & 98.56 & 13.08 &  (24) & 46 & 16 & 2.88x \\ 
  2D Conv. & 60.31 & 10.18 &  (24) & 66 & 33 & 2.00x \\ 
  Reduction & 28.31 & 4.21 &  (16) & 43 & 11 & 3.91x \\ 
  Histogram & 11.86 & 7.53 &  (24) & 61 & 8 & 7.62x \\ 
  Sparse Mult. & 2.85 & 0.63 & (20) & 51 & 14 & 3.64x \\ 
  Black Scholes  & 5.93 & \multicolumn{2}{c}{-}  & - & - & - \\ 
 Cor. Matrix & 26.16 & \multicolumn{2}{c}{-}  & - & - & - \\ 
  \midrule
   Mean & 31.94 & 6.94 && 51 & 15 & 4.45x \\ 
   \bottomrule
\end{tabular}
\caption{}
\label{tbl:results_summary}
\end{subtable}
\caption{(Left)Speedup obtained by APARAPI and Jacc over serial Java implementations. (Right) A comparison of Jacc against Java based implementations of the benchmarks. (Thread counts are in brackets). }
\end{figure*}

The two metrics used for evaluating Jacc are raw performance and programmability. Regarding raw performance, Jacc is thoroughly evaluated against: 1) serial Java implementations, 2) multithreaded Java implementations,  3) multithreaded OpenMP implementations, and 4) alternative GPGPU offloading techniques used in the APARAPI \cite{aparapi}. The effect on programmability is studied by comparing the lines of code required to implement Jacc GPGPU accelerated code against vanilla multithreaded Java code.


Table \ref{tbl:results_summary} summarizes the speedups obtained by Jacc only against Java implementations (Speedup column).
The speedups are normalized to the performance of two different Java implementations: a serial Java implementation and the peak performing multi-threaded Java implementation.
Results indicate that Jacc constantly outperforms all Java implementations both serial (32x average) and best multithreaded (7x average) with the exception of sparse vector multiplication. 
The irregular memory access pattern (presence of lookup tables hindering the ahead-of-time balancing) of sparse vector multiplication does not favor GPGPU execution. 

\subsection{Heterogeneous Code Size}

To evaluate the programmability benefits of Jacc in comparison to the multithreaded implementations we compare the code sizes of both of them.
Table \ref{tbl:results_summary} provides the lines of code required to create the parallel kernel for each benchmark.
The results show that on average a Jacc implementation requires 4.45x less code.

\subsection{APARAPI Comparison}

Our comparison against APARAPI is based on three benchmarks: one benchmark (vector addition) and two provided by APARAPI (Black Scholes and Correlation Matrix).

Figure \ref{fig:aparapi_comparison} compares the speedups achieved by Jacc and APAR\-API across these benchmarks.
To understand the impact of JIT compilation on performance, we include performance data for experiments that are both inclusive and exclusive of compilation time.
Comparing the geometric mean of these speedups, we observe that both frameworks are very similar in terms of performance; APARAPI is better if compilation times are included and Jacc is better if compilation times are excluded.

In constrast to our approach, APARAPI is built upon OpenCL and uses source-to-source translation to generate OpenCL C from Java bytecode.
This approach provides APARAPI with two advantages: consistently low-compilation times, around 400 milliseconds, and a high quality of generated code.
In Figure \ref{fig:aparapi_comparison}, this manifests as smaller differences between the performance data that includes and excludes compilation times.
Conversely, Jacc has a larger performance gap when including and excluding compilation times - something that will shrink as our compiler matures.

In the Correlation Matrix benchmark, Jacc significantly outperforms APARAPI.
This is due to two factors: the ability to easily tune the number of threads in each work group\footnote{We found that changing Jacc's work group size, to match that of APARAPI, severely reduced performance but remained faster than APARAPI.} and the ability to take advantage of the GPGPU's \texttt{popc} instruction.

\section{Related Work}
\label{sec:related_work}

Currently, the related work spans into two areas: 1) providing support for heterogeneous offloading (specifically in the Java context), and 2) improving heterogeneous programming in the general case - typically using a C based language.

The OpenJDK Sumatra project \cite{project_sumatra} currently represents the state of the art in integrating GPGPU offload in Java.
Sumatra aims to bring GPGPU offload mainstream with the help of Java 8 features such as lambda functions and Streams. Thus, Sumatra does not help with previous versions of Java and its GPGPU offload is heavily dependent on developers using Java 8 feautres. Furthermore, the project incorporates knowledge gained from earlier projects, such as AMD's APARAPI \cite{aparapi} and Rootbeer \cite{rootbeer}. Sumatra differs from Jacc because it mainly focuses on providing a framework for GPU acceleration. Jacc, on the other hand, is a more generic framework for accelerating Java on multiple hardware accelerators due to its modular design. Additionally, Jacc provides the flexibility to conditionally schedule offloads depending on hardware availability or on problem size.

Habanero-Java \cite{habanero_java} provides the ability to offload Java applications onto hardware accelerators using OpenCL \cite{OpenCL}. Jacc differentiates in terms of language integration and code generation. While Habanero-Java requires developers to write code using the HJ language, which is in turn compiled to OpenCL C code, Jacc can convert bytecode into lower level assembly or machine code, such as PTX \cite{ptx}.

Prior to the Sumatra  project, a number of other projects such as JCuda \cite{jcuda_2009}, JCudaMP \cite{jcudamp_2010} and the Java OpenCL bindings \cite{jocl} implemented support for GPGPU offload. Jacc differentiates from those approaches since it does not employ source-to-source translation. In contrast, the implemented JIT compiler directly translates Java bytecodes to native code for accelerator offloading.

Outside the Java world, CUDA \cite{cuda}, OpenAcc \cite{openacc} and OpenCL \cite{OpenCL} are established programming models for heterogeneous computing.
While CUDA is exclusively focusing on programming GPGPUs, OpenAcc and OpenCL support a wider variety of hardware. Since these programming models are primarily focusing on assisting developers in maximizing the performance of their applications, they allow low level access to hardware. The low level hardware access enables expert developers to write high performance code, whilst creating a steep learning curve for inexperienced developers.

\section{Conclusions}
\label{conclusions}

Heterogeneous programming allows developers to improve performance by running portions of code on the most appropriate hardware resource. 
In this paper we have introduced Jacc, a framework which allows Java to be used for heterogeneous programming. 
Moreover, we have showcased the ability of Jacc to utilise GPGPU accelerators.
Our experimental results demonstrate the advantages of Jacc, both in terms of programmability and performance, by evaluating it against existing Java frameworks for multi-threaded programming and heterogeneous programming.  
Experimental results show an average performance speedup of 32x and a 4.4x code decrease across eight evaluated benchmarks on a NVIDIA Tesla K20m GPU.

In the future, our intent is to build on this foundation and increase our coverage of the Java language.
One major obstacle is integrating heterogeneous offload with the JVM, and more specifically the garbage collector to allow objects to be created on heterogeneous devices.
Furthermore, we are planning to extend Jacc to support a range of heterogeneous devices, such as FPGAs and multi-core processors.

\acks
This work is supported by the AnyScale Apps and PAMELA projects funded by EPSRC EP/L000725/1 and EP/K008730/1.
Dr Luj{\'a}n is supported by a Royal Society University Research Fellowship.

\bibliographystyle{abbrvnat}
\bibliography{references}

\end{document}